\documentclass[12pt]{article}

\usepackage[dvips]{graphicx}
\usepackage[english]{babel}

\usepackage{amsmath}

\allowdisplaybreaks[2]

\newcommand{\half}{{\textstyle\frac{1}{2}}}

\newcommand{\gev}{\operatorname{GeV}}

\newcommand{\nb}{\operatorname{nb}}

\newcommand{\ms}{\mskip 1.5mu}

\newcommand{\gsim}{\raisebox{-4pt}{%
    $\;\stackrel{\textstyle >}{\sim}\;$}}

\newcommand{\KK}{K^0\smash{\overline{K}{}^0}}
\newcommand{\MM}{M\smash{\overline{M}}}
\newcommand{\qq}{q\smash{\overline{q}}}

\newcommand{\tw}{\textwidth}

\begin{document}
\thispagestyle{empty}
\begin{flushright}
DESY 09-204 \\
WU B 09-17
\end{flushright}

\begin{center}
\vskip 4.0\baselineskip
\textbf{\LARGE Two-photon annihilation into \\[0.1em]
  octet meson pairs: \\[0.3em]
  symmetry relations in the handbag approach}

\vskip 8.0\baselineskip
M.~Diehl \\[0.5\baselineskip]
\textit{Deutsches Elektronen-Synchroton DESY, 22603 Hamburg, Germany}
\\[1.5\baselineskip]
P.~Kroll \\[0.5\baselineskip]
\textit{Fachbereich Physik, Universit\"at Wuppertal, 42097 Wuppertal,
  Germany \\
  and \\
  Institut f\"ur Theoretische Physik, Universit\"at Regensburg,
  93040 Regensburg, Germany}
\vskip 4.0\baselineskip
\textbf{Abstract}\\[0.8\baselineskip]
\parbox{0.9\textwidth}{We explore the implications of SU(3) flavor
  symmetry in the soft handbag mechanism for two-photon annihilation into
  pairs of pseudoscalar octet mesons.  In this approach we obtain a good
  description of the experimental results for all measured channels at
  high energy, with two complex form factors adjusted to the data.  We
  also predict the cross section for $\gamma\gamma\to \eta\eta$.}
\end{center}

\vfill

\section{Introduction}
\label{sec:1}

A few years ago, we have investigated two-photon annihilation into pairs
of pseudoscalar mesons within the soft handbag approach
\cite{Diehl:2001fv}.  The physical mechanism described by this approach is
a parton-level process $\gamma\gamma\to \qq$ followed by hadronization of
the $q\bar{q}$ pair into the final-state mesons, where each meson
approximately carries the momentum of the quark or antiquark.  To ensure
that the parton-level process proceeds at short distances, one must
require large Mandelstam variables $s$, $t$, and $u$.  Neglecting
contributions that are suppressed in this kinematics, we have derived the
cross section formula
\begin{equation}
 \frac{d\sigma}{d t}(\gamma\gamma \to \MM) = 
    \frac{8 \pi \alpha_{\text{em}}^2}{s^2} \, \frac{1}{\sin^4\theta}\;
    \big| R_{\MM}(s) \big|^{2} ,
\label{eq:dsdt-MM}
\end{equation}
where $\theta$ is the c.m.\ scattering angle.  The form factor
$R_{\MM}(s)$ describes the soft transition from $\qq$ to the meson pair.
It can be written as the matrix element of the quark part of the
energy-momentum tensor, summed over quark flavors with appropriate charge
factors and taken between the $\MM$ state and the vacuum.  A key
prediction of the handbag approach is the absence of final states with
isospin $I=2$, given that the process proceeds through a single
intermediate $\qq$ pair.  Using approximate flavor SU(3) symmetry for the
transition form factors one obtains a number of further predictions, which
are discussed in detail in the present work.

The soft handbag mechanism for $\gamma\gamma$ annihilation into a hadron
pair has a spacelike analog in Compton scattering at large Mandelstam
variables \cite{Radyushkin:1998rt}, and its dynamics is closely related
with the Feynman mechanism for form factors at large momentum transfer.
As in the latter case, one can show that at asymptotically large momenta
another mechanism will be dominant, namely the hard-scattering mechanism
of Brodsky and Lepage \cite{Lepage:1980fj}.  For the process under
discussion, the amplitude then factorizes into a hard partonic subprocess
$\gamma\gamma\to \qq\qq$ and a single-meson distribution amplitude for
each meson \cite{Brodsky:1981rp}.  Since there are two $\qq$ pairs in the
intermediate state, $I=2$ transitions are allowed, in contrast to the
handbag approach.  Which mechanism is relevant in given kinematics can of
course not be determined by general arguments, and one of the purposes of
this paper is a detailed comparison of soft handbag results with data.

In the past years, the BELLE collaboration has published data on exclusive
$\gamma\gamma$ annihilation for many different pseudoscalar meson channels
\cite{Nakazawa:2004gu,Uehara:2009cka,Chen:2006gy,Uehara:2009cf}.  These
data are significantly more accurate and cover more final states than
previous measurements \cite{Dominick:1994bw,Heister:2003ae}, which were
only available for $\pi^+\pi^-$ and $K^+K^-$ pairs.  In view of this
situation we find it timely to update our initial analysis in
\cite{Diehl:2001fv}.  The BELLE data
\cite{Nakazawa:2004gu,Uehara:2009cka,Chen:2006gy,Uehara:2009cf} nicely
confirm the $1/\sin^4{\theta}$ behavior \eqref{eq:dsdt-MM} of the
differential cross sections for $s$ larger than about $9 \gev^2$.  This
allows us to determine the process form factors from cross sections
integrated over a given interval of $\theta$ around $90^\circ$, where the
handbag approach is applicable.  The cross section integrated over
$\cos\theta$ from $-\cos\theta_0$ to $\cos\theta_0$ reads
\begin{equation}
\sigma(\gamma\gamma \to \MM) = S\, \frac{4 \pi \alpha_{\text{em}}^2}{s}
  \biggl[\ms \frac{\cos\theta_0}{\sin^2\theta_0} 
    + \frac{1}{2} \ln\frac{1+\cos\theta_0}{1-\cos\theta_0} \ms\biggr] \,
 \big| R_{\MM}(s) \big|^{2} ,
\label{eq:sig}
\end{equation}
where the statistical factor $S$ is equal to $1/2$ for two identical
mesons like $\pi^0\pi^0$ and equal to unity otherwise.  Most of the
experimental data for the integrated cross section are quoted for
$\cos\theta_0=0.6$.  An exception is the $\eta\pi^0$ cross section in
\cite{Uehara:2009cf}, which is given for $\cos\theta_0=0.8$.  The numerical
constants that give the cross section \eqref{eq:sig} when multiplied with
$s^{-1}\, |R_{\MM}(s)|^2$ are $425 \nb \gev^2$ for $\cos\theta_0=0.6$ and
$866 \nb \gev^2$ for $\cos\theta_0=0.8$ in the case where $S=1$.

The plan of this paper is as follows.  In the next section we discuss the
flavor-symmetry properties of the handbag amplitude, with particular
emphasis on the $\eta\pi^0$ and $\eta\eta$ channels, which have not been
considered before and add to the predictive power of the approach.  In
Sect.~\ref{sec:pion} we extract the annihilation form factor $R_{\MM}$ for
each of the available meson channels, while in Sect.~\ref{sec:combined} we
analyze the annihilation form factors in terms of the two independent
quark-level form factors, a valence and a non-valence one.  In
Sect.~\ref{sec:brodsky-lepage} we point out some differences between the
handbag approach and the hard-scattering mechanism of Brodsky and Lepage,
before summarizing our main results in Sect.~\ref{sec:sum}.

\section{Flavor symmetry relations in the handbag approach}
\label{sec:symmetry}

To derive flavor SU(3) relations between the annihilation amplitudes into
different meson pairs, we find it convenient to consider the usual
isospin multiplets and in addition the multiplets under $U$-spin and
$V$-spin, which are the SU(2) symmetries associated with the exchange $s
\leftrightarrow d$ and $u \leftrightarrow s$, respectively
\cite{London:1964zz}.  The corresponding singlets, doublets, and triplets
read
\begin{align}
U: & \qquad \half \bigl( \eta + \sqrt{3}\pi^0 \bigr) \,,
     \qquad \bigl\{ \pi^+ ,K^+ \bigr\} \,,
     \qquad \bigl\{ \overline{K}{}^0,
            \half \bigl( \sqrt{3}\eta - \pi^0 \bigr), K^0 \bigr\} \,,
\nonumber \\
V: & \qquad \half \bigl( \eta - \sqrt{3}\pi^0 \bigr) \,,
     \qquad \bigl\{ K^0, \pi^- \bigr\} \,,
     \qquad \bigl\{ K^+, \half \bigl( \sqrt{3}\eta + \pi^0 \bigr),
                    K^- \bigr\} \,,
\end{align}
where our phase conventions for meson states correspond to the quark
content
\begin{align}
\pi^+ &= \bar{d}u \,, & 
\pi^- &= \bar{u}d \,, &
\pi^0 &= \tfrac{1}{\sqrt{2}}\ms \bigl( \bar{u}u - \bar{d}d \bigr) \,,
\nonumber \\
K^+ &= \bar{s}u \,, & 
K^- &= \bar{u}s \,, \phantom{\tfrac{1}{\sqrt{2}}} 
\nonumber \\
K^0 &= \bar{s}d \,, & 
\overline{K}{}^0 &= \bar{d}s \,, &
\eta &= \tfrac{1}{\sqrt{6}}\ms 
        \bigl( \bar{u}u + \bar{d}d - 2\bar{s}s \bigr) \,.
\end{align}
At this stage we neglect $\eta$-$\eta^\prime$ mixing and approximate the
physical $\eta$ as a pure flavor octet state.
In the process amplitudes and annihilation form factors for $\gamma\gamma
\to M_1 M_2$, the two-meson states appear in the $C$-even combinations
\begin{equation}
  \label{pair-states}
|M_1 M_2\rangle = \frac{\big| M_1(p) M_2(p') \big\rangle
                      + \big| M_1(p') M_2(p) \big\rangle}{2} \,,
\end{equation}
so that $|M_1 M_2\rangle = |M_2 M_1\rangle$.  {}From these two-meson
states we can form linear combinations $|\Phi_{I} \rangle$, $|\Phi_{U}
\rangle$, $|\Phi_{V} \rangle$ with definite isospin, $U$-spin, and
$V$-spin, respectively.

$U$-spin symmetry plays a special role in our context, because the photon
is a $U$-spin singlet.  From the absence of two-photon transitions to the
states $\Phi_{U=1}$ and $\Phi_{U=2}$ we immediately obtain the following
relations for the amplitudes of $\gamma\gamma \to M_1 M_2$\,:
\begin{align}
  \label{rel-SU3-1}
\mathcal{A}_{K^+K^-} &= \mathcal{A}_{\pi^+\pi^-} \,,
\\
  \label{rel-SU3-2}
\mathcal{A}_{\KK} &= \tfrac{3}{4} \mathcal{A}_{\eta\eta}
  + \tfrac{1}{4} \mathcal{A}_{\pi^0\pi^0}
  - \tfrac{\sqrt{3}}{2} \mathcal{A}_{\eta\pi^0} \,.
\end{align}
These relations are respected by \emph{any} dynamical mechanism in the
limit of SU(3) symmetry.

A further set of relations emerges in the soft handbag mechanism because
the two photons annihilate via a quark-antiquark intermediate state.  This
allows only final states with $I$ and $V$ equal to $0$ or $1$.  The
absence of $I=2$ and $V=2$ states respectively implies
\begin{align}
  \label{rel-haba-1}
\mathcal{A}_{\pi^0\pi^0} &= \mathcal{A}_{\pi^+\pi^-} \,,
\\
  \label{rel-haba-2}
\mathcal{A}_{K^+K^-} &= \tfrac{3}{4} \mathcal{A}_{\eta\eta}
  + \tfrac{1}{4} \mathcal{A}_{\pi^0\pi^0}
  + \tfrac{\sqrt{3}}{2} \mathcal{A}_{\eta\pi^0} \,.
\end{align}
In the SU(3) limit, the handbag mechanism thus predicts equal differential
cross sections for the channels $K^+K^-$, $\pi^+\pi^-$, and $\pi^0\pi^0$,
as was already pointed out in \cite{Diehl:2001fv}.  With this equality,
the sum and the difference of \eqref{rel-SU3-2} and \eqref{rel-haba-2}
give relations between only three different channels:
\begin{align}
  \label{rel-pair-1}
\mathcal{A}_{\eta\pi^0} &= 
\tfrac{1}{\sqrt{3}}\ms
   \bigl( \mathcal{A}_{K^+K^-} - \mathcal{A}_{\KK} \bigr) \,,
\\
  \label{rel-pair-2}
\mathcal{A}_{\eta\eta} &= \tfrac{1}{3}\ms \mathcal{A}_{K^+K^-}
     + \tfrac{2}{3}\ms \mathcal{A}_{\KK} \,,
     \phantom{\tfrac{1}{\sqrt{3}}}
\intertext{or equivalently}
  \label{rel-pair-3}
\mathcal{A}_{K^+K^-}
  &= \mathcal{A}_{\eta\eta}
   + \tfrac{2}{\sqrt{3}}\ms \mathcal{A}_{\eta\pi^0} \,,
\\
  \label{rel-pair-4}
\mathcal{A}_{\KK}
  &= \mathcal{A}_{\eta\eta} 
   - \tfrac{1}{\sqrt{3}}\ms \mathcal{A}_{\eta\pi^0} \,.
\end{align}
Let us work out the consequences of \eqref{rel-pair-1} to
\eqref{rel-pair-4}, which involve the final states $K^+K^-$, $\KK$,
$\eta\pi^0$, and $\eta\eta$.  The square of each relation contains a term
depending on the relative phase between the two amplitudes on the
right-hand side.  This yields triangular inequalities between any
combination of three cross sections.  The first of these reads
\begin{equation}
\Biggl|\, \sqrt{\frac{1}{3}\ms\frac{d\sigma_{K^+K^-}}{dt}}
 - \sqrt{\frac{1}{3}\ms\frac{d\sigma_{\KK}}{dt}} \,\Biggr| \;\le\;
\sqrt{\frac{d\sigma_{\eta\pi^0}}{dt}} \;\le\;
   \sqrt{\frac{1}{3}\ms\frac{d\sigma_{K^+K^-}}{dt}}
 + \sqrt{\frac{1}{3}\ms\frac{d\sigma_{\KK}}{dt}} \,,
\label{eq:bound}
\end{equation}
and the others are readily obtained in analogy.  One obtains one relation
between all four cross sections in which phases between amplitudes drop
out:
\begin{equation}
2\ms \frac{d\sigma_{\eta\pi^0}}{dt} + 3\ms \frac{d\sigma_{\eta\eta}}{dt}
 =  \frac{d\sigma_{K^+K^-}}{dt}
    + 2\ms \frac{d\sigma_{\KK}}{dt} \,,
\label{eq:cross-section-relation}
\end{equation}
which may be regarded as the SU(3) analog of the isospin relation
$d\sigma_{\pi^0\pi^0} /dt = d\sigma_{\pi^+\pi^-} /dt$.

Having derived relations between the full process amplitudes in the
handbag mechanism, we now turn our attention to the annihilation form
factors for each separate quark flavor.  SU(3) flavor symmetry gives
\begin{align}
R^u_{\Phi_{I=0}} &= R^d_{\Phi_{I=0}} \,, &
R^u_{\Phi_{I=1}} &= - R^d_{\Phi_{I=1}} \,, &
R^s_{\Phi_{I=1}} &= R^s_{\Phi_{I=2}}
                  = R^u_{\Phi_{I=2}} = R^d_{\Phi_{I=2}} = 0 \,,
\nonumber \\
R^s_{\Phi_{U=0}} &= R^d_{\Phi_{U=0}} \,, &
R^s_{\Phi_{U=1}} &= - R^d_{\Phi_{U=1}} \,, &
R^u_{\Phi_{U=1}} &= R^u_{\Phi_{U=2}}
                  = R^s_{\Phi_{U=2}} = R^d_{\Phi_{U=2}} = 0
\end{align}
for the annihilation form factors into states with definite isospin or
$U$-spin.  With this we can express all annihilation form factors in terms
of a valence form factor
\begin{align}
  \label{val-form-fact}
R^u_{2\pi} &\stackrel{\text{def}}{=} R^u_{\pi^+\pi^-}
  = R^d_{\pi^+\pi^-} = R^u_{\pi^0\pi^0} = R^d_{\pi^0\pi^0} \,,
\nonumber \\
 &= R^{u}_{K^+K^-} = R^{s}_{K^+K^-}
  = R^{d}_{\KK} = R^{s}_{\KK}
\end{align}
and a non-valence form factor
\begin{align}
R^s_{2\pi} &\stackrel{\text{def}}{=} R^s_{\pi^+\pi^-}
  = R^s_{\pi^0\pi^0}
  = R^{d}_{K^+K^-} = R^{u}_{\KK} \,.
\end{align}
These results were already given in \cite{Diehl:2001fv}.  In addition we
obtain
\begin{align}
  \label{quark-relations}
R^{u}_{\eta\pi^0} &= - R^{d}_{\eta\pi^0}
  = \tfrac{1}{\sqrt{3}} \bigl( R^{u}_{2\pi} - R^{s}_{2\pi} \bigr) \,,
&
R^{s}_{\eta\pi^0} &= 0 \,,
\nonumber \\
R^{u}_{\eta\eta} &= \phantom{-} R^{d}_{\eta\eta^{\phantom{0}}}
  = \tfrac{1}{3} R^{u}_{2\pi} + \tfrac{2}{3} R^{s}_{2\pi} \,,
&
R^{s}_{\eta\eta} &= \tfrac{4}{3} R^{u}_{2\pi} - \tfrac{1}{3} R^{s}_{2\pi} \,.
\end{align}
It is instructive to compare these results with the corresponding flavor
SU(3) relations for the production of $B\overline{B}$ pairs, where $B$ is
a member of the ground state baryon octet \cite{Diehl:2002yh}.  In the
meson case there are two independent form factors, whereas in the baryon
case there are three.  This is because there are fewer meson-antimeson
states, given that the antimeson octet is equal to the meson octet.  In
the baryon case we have for instance two distinct states $|\Sigma^+
\overline{\Sigma}{}^-\rangle$ and $|\Sigma^-
\overline{\Sigma}{}^+\rangle$, whereas the corresponding meson states
$|\pi^+ \pi^-\rangle$ and $|\pi^- \pi^+\rangle$ are equivalent according
to \eqref{pair-states}.  Taking this into account, the relations
\eqref{val-form-fact} to \eqref{quark-relations} are the analog of
eq.~(40) in \cite{Diehl:2002yh}.

Summing up the individual quark flavor contributions to each channel as
\begin{equation}
R_{\MM}^{} =
\tfrac{4}{9} R_{\MM}^u + \tfrac{1}{9} R_{\MM}^d + \tfrac{1}{9} R_{\MM}^s \,,
\end{equation}
we find that \eqref{val-form-fact} to \eqref{quark-relations} implies
\begin{equation}
  \label{old-form-factors}
R_{\pi^0\pi^0} = R_{\pi^+\pi^-} = R_{K^+K^-}
\end{equation}
and
\begin{align}
  \label{form-factors}
R_{\pi^+\pi^-}
  &= \tfrac{5}{9} R^{u}_{2\pi} + \tfrac{1}{9} R^{s}_{2\pi} \,,
  \phantom{\tfrac{1}{\sqrt{3}}}
\nonumber \\
R_{\KK}
  &= \tfrac{2}{9} R^{u}_{2\pi} + \tfrac{4}{9} R^{s}_{2\pi} \,,
  \phantom{\tfrac{1}{\sqrt{3}}}
\nonumber \\
R_{\eta\pi^0}
  &= \tfrac{1}{3\sqrt{3}} \bigl( R^{u}_{2\pi} - R^{s}_{2\pi} \bigr) \,,
\nonumber \\
R_{\eta\eta}
  &= \tfrac{1}{3} \bigl( R^{u}_{2\pi} + R^{s}_{2\pi} \bigr) \,.
\end{align}
Only the equality $R_{\pi^+\pi^-} = R_{\pi^0\pi^0}$ and the first relation
in \eqref{form-factors} follow from isospin symmetry alone, whereas all
other relations need full SU(3) flavor symmetry.  {}From
\eqref{form-factors} one can readily rederive the relations
\eqref{rel-pair-1} to \eqref{rel-pair-4} for the process amplitudes.

If one assumes that the soft handbag mechanism is valid within a certain
accuracy, one can use the cross sections for any three of the channels
$K^+K^-$, $\KK$, $\eta\pi^0$, $\eta\eta$ to extract the size of the form
factors $R_{2\pi}^{u}$ and $R_{2\pi}^{s}$ as well as their relative phase
$\rho$.  Taking the squares of the first three relations in
\eqref{form-factors} and forming appropriate linear combinations, we
obtain for instance
\begin{align}
f(s,\theta) \, \big| R^{u}_{2\pi} \big|^2
 &=   -  \frac{d\sigma_{\KK}}{dt}
   + 4\, \frac{d\sigma_{\eta\pi^0}}{dt}
   + 4\, \frac{d\sigma_{K^+K^-}}{dt} \,,
\nonumber \\
f(s,\theta) \, \big| R^{s}_{2\pi} \big|^2
 &=  5\, \frac{d\sigma_{\KK}}{dt}
  + 10\, \frac{d\sigma_{\eta\pi^0}}{dt}
   - 2\, \frac{d\sigma_{K^+K^-}}{dt} \,,
\nonumber \\
f(s,\theta) \, \big| R^{u}_{2\pi} R^{s}_{2\pi} \big| \cos\rho
 &=  2\, \frac{d\sigma_{\KK}}{dt}
  - 11\, \frac{d\sigma_{\eta\pi^0}}{dt}
       + \frac{d\sigma_{K^+K^-}}{dt} \,,
\label{eq:basic}
\end{align}
where
\begin{equation}
f(s,\theta) = \frac{32\pi \alpha_{\text{em}}^2}{3 s^2 \sin^4\theta} \,.
\end{equation}

\section{Extraction of the process form factors}
\label{sec:pion}

\begin{figure}[ht]
\begin{center} 
\includegraphics[width=.45\tw,bb=117 282 486 693,clip=true]
{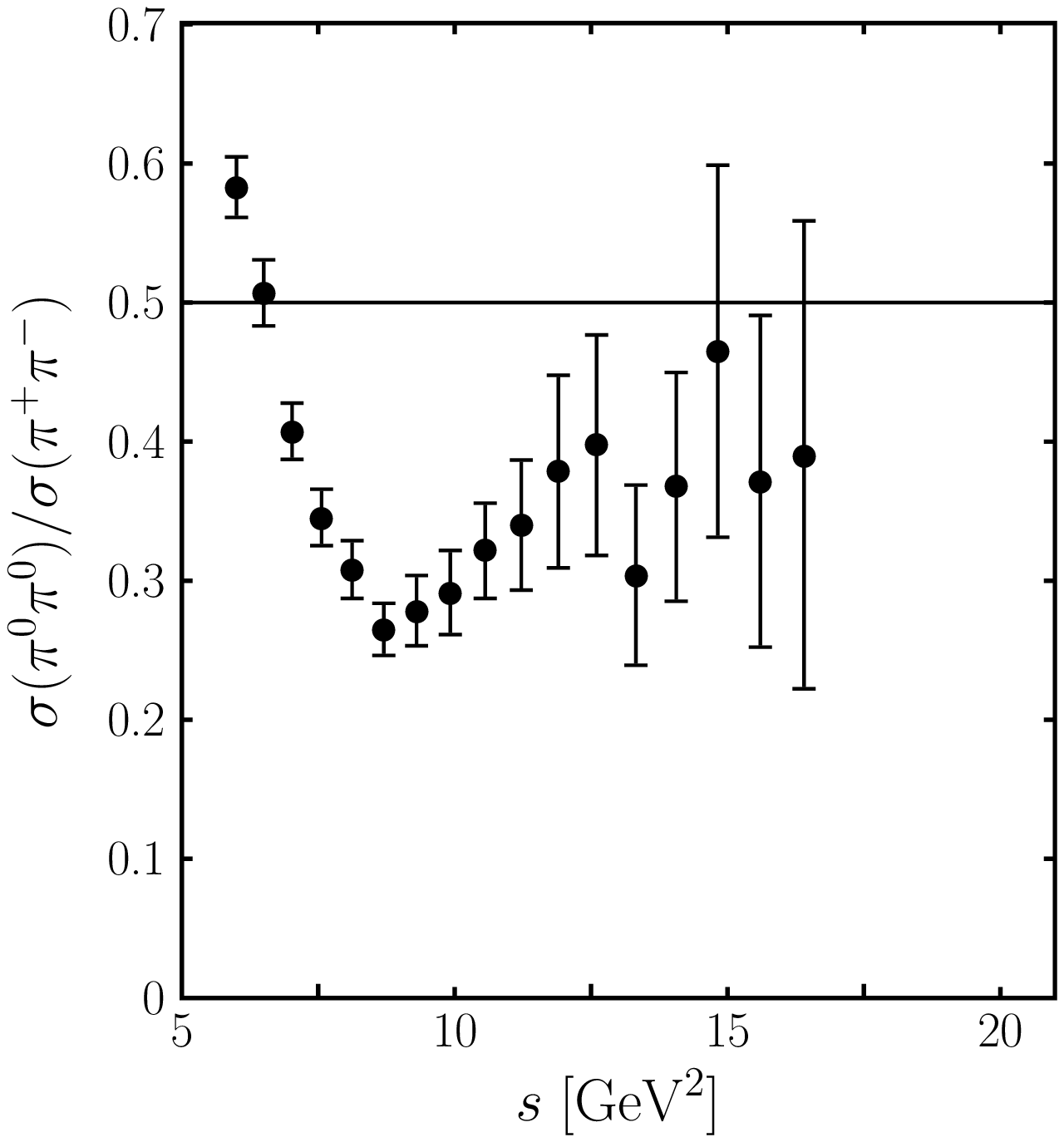}
\end{center}
\caption{\label{fig:cross-ratio} BELLE data \protect\cite{Uehara:2009cka}
  for the cross section ratio of neutral and charged pion pair production.
  Contributions from charmonium decay have been subtracted by BELLE.  The
  errors shown are statistical only.}
\end{figure}

The amplitudes for the production of pion pairs in two-photon annihilation
have the isospin decomposition
\begin{align}
  \label{eq:isospin-decomp}
\mathcal{A}_{\pi^+\pi^-} &= 
   \tfrac{1}{\sqrt{3}}\, \mathcal{A}_{2\pi}^{I=0}
 + \tfrac{1}{\sqrt{6}}\, \mathcal{A}_{2\pi}^{I=2}\,,
\nonumber \\
\mathcal{A}_{\pi^0\pi^0} &= 
   \tfrac{1}{\sqrt{3}}\, \mathcal{A}_{2\pi}^{I=0}
 - \tfrac{2}{\sqrt{6}}\, \mathcal{A}_{2\pi}^{I=2}\,,
\end{align}
with the $I=1$ final state being forbidden by charge conjugation.  The
absence of $I=2$ transitions, which is a central property of the handbag
approach as discussed above, gives
\begin{equation}
R \,\stackrel{\text{def}}{=}\,
\frac{\sigma(\gamma\gamma\to \pi^0\pi^0)}{\sigma(\gamma\gamma\to \pi^+\pi^-)}
  = \frac{1}{2}
\label{eq:ratio}
\end{equation}
for the ratio of integrated cross sections, where the statistical factor
$S=1/2$ is taken into account for the $\pi^0\pi^0$ channel.  By contrast,
one has $R=2$ if $I=0$ transitions are absent.  As shown in
Fig.~\ref{fig:cross-ratio}, the experimental value of $R$ is about 0.3 at
$s=9\,\gev^2$ and increases for higher energies, with central values
between 0.3 and 0.5 and rapidly growing errors.

Because amplitudes enter quadratically in cross sections, one cannot
uniquely reconstruct the ratio $r=\mathcal{A}_{2\pi}^{I=2}
/\mathcal{A}_{2\pi}^{I=0}$ of isospin amplitudes from the cross section
ratio $R$, even for a given relative phase between
$\mathcal{A}_{2\pi}^{I=0}$ and $\mathcal{A}_{2\pi}^{I=2}$.  For instance,
both $r=0$ and $r=2\sqrt{2}$ give $R = 1/2$, and both $r=\sqrt{2}/14
\approx 0.1$ and $r=3/\sqrt{2} \approx 2.1$ give $R =8/25 =0.32$.  If one
\emph{assumes} that the two amplitudes are in phase and that
$\mathcal{A}_{2\pi}^{I=2} < \mathcal{A}_{2\pi}^{I=0}$, then $0.3 < R <
0.5$ implies $\mathcal{A}_{2\pi}^{I=2}/ \mathcal{A}_{2\pi}^{I=0} < 0.11$.
We thus see that the data on $\pi^0\pi^0$ and $\pi^+\pi^-$ production for
$s > 9 \gev^2$ cannot prove the dominance of $I=0$ transitions but is
fully consistent with it.

\begin{figure}
\begin{center} 
\includegraphics[width=.45\tw,bb=117 262 486 693,clip=true]
{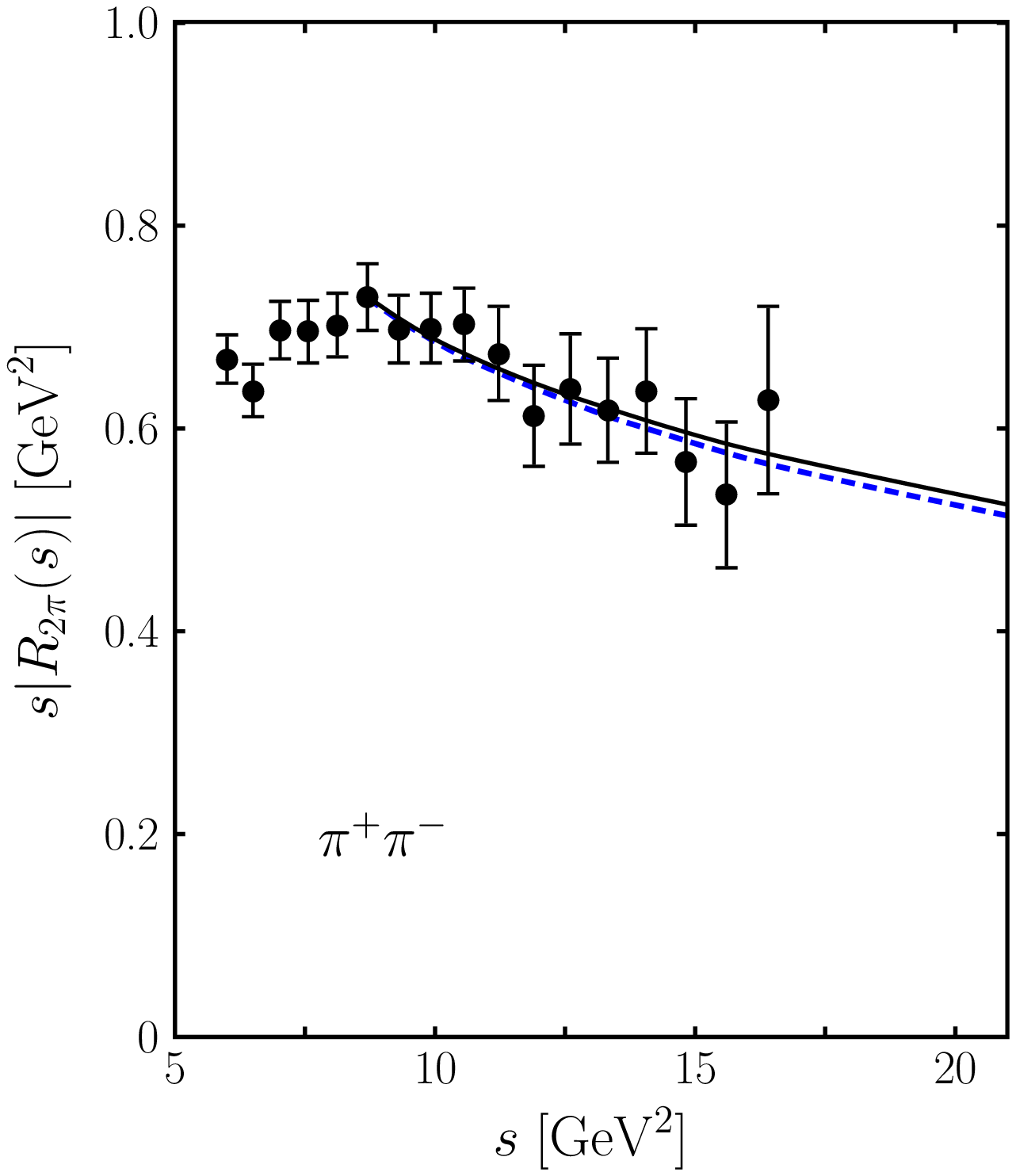}
\includegraphics[width=.49\tw,bb=125 256 534 695,clip=true]
{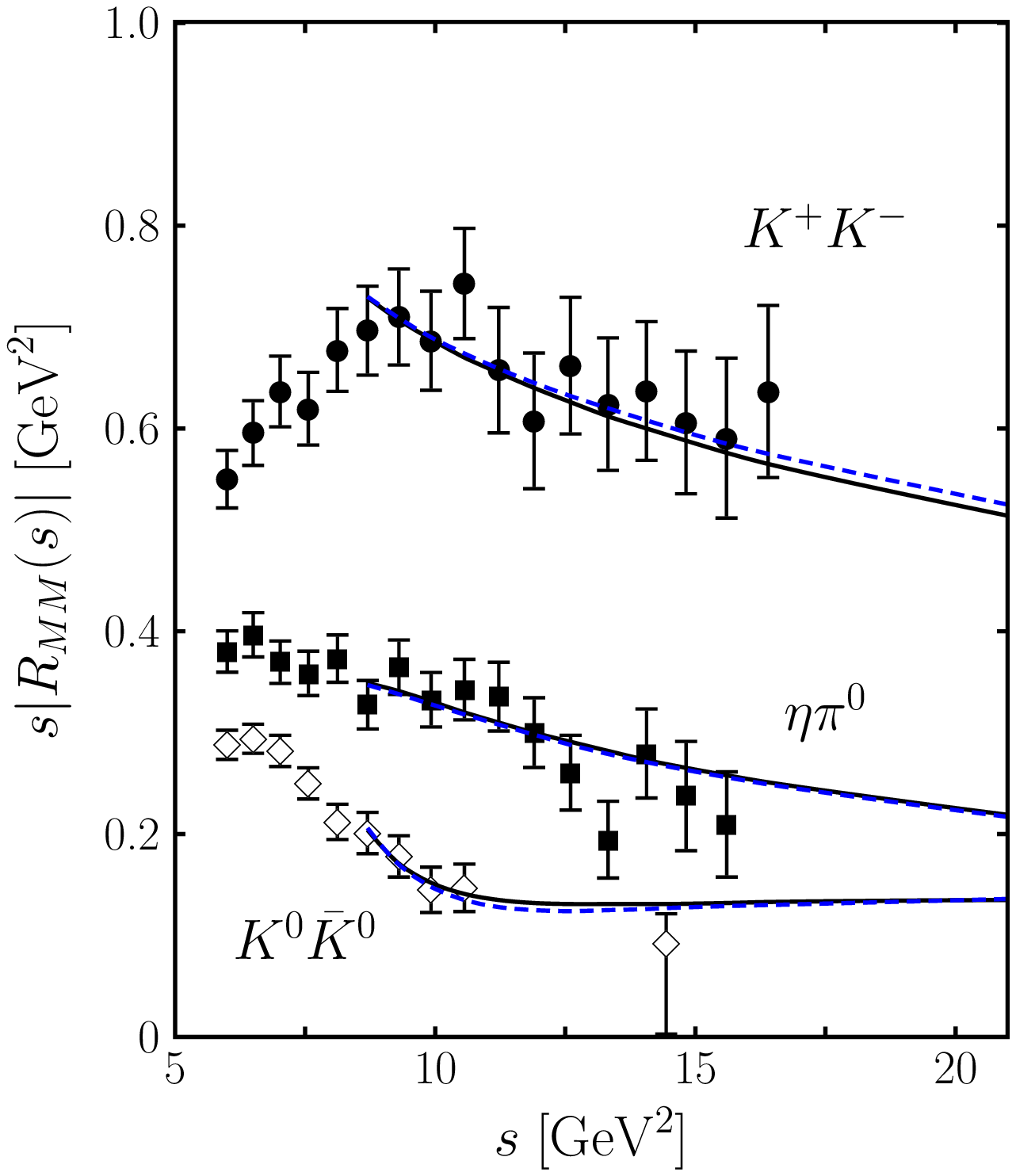}
\end{center}
\caption{\label{fig:FFpipi} \emph{Left:} The modulus of the scaled
  annihilation form factor $s R_{\pi\pi}(s)$, extracted from the
  combination \protect\eqref{eq:i=0} of $\pi^0\pi^0$ and $\pi^+\pi^-$
  cross sections.  Data are taken from
  \protect\protect\cite{Nakazawa:2004gu,Uehara:2009cka}.  \emph{Right:}
  The scaled annihilation form factors for the $K^+K^-$, $K^0\bar{K}^0$,
  and $\eta\pi^0$ channels.  Data are from \protect\cite{Nakazawa:2004gu}
  for $K^+K^-$ (filled circles), from \protect\cite{Uehara:2009cf} for
  $\eta\pi^0$ (filled squares), and from \cite{Chen:2006gy} for
  $K^0\bar{K}^0$ (open diamonds).  Here and in all subsequent plots, the
  error bars combine statistical and systematic uncertainties in
  quadrature, and solid (dashed) lines represent fit 1 (2) specified in
  Tab.~\protect\ref{tab:2}.}
\end{figure}

With this situation in mind we determine the process form factor
$R_{\pi\pi}$ from the combination
\begin{equation}
\frac{2}{3} \Big[ \sigma(\gamma\gamma\to \pi^0\pi^0)+
                  \sigma(\gamma\gamma\to\pi^+\pi^-) \Big] \,,
\label{eq:i=0} 
\end{equation}
where the interference term between the $I=0$ and $I=2$ transitions
cancels and where the squared amplitudes $| \mathcal{A}_{2\pi}^{I=0} |^2$
and $| \mathcal{A}_{2\pi}^{I=2} |^2$ enter with equal weight.  The form
factor obtained from this combination is almost pure $I=0$ and only mildly
contaminated by the $I=2$ contribution, provided that
$|\mathcal{A}_{2\pi}^{I=2}/ \mathcal{A}_{2\pi}^{I=0}|$ is small compared
to 1, which we assume from now on.  The results for $R_{\pi\pi}(s)$
obtained from the BELLE data \cite{Nakazawa:2004gu,Uehara:2009cka} are
shown in the left panel of Fig.~\ref{fig:FFpipi}.  Here and in the
following we add statistical and systematic errors of the data in
quadrature.

We proceed by extracting the process form factors for the $K^+K^-$, $\KK$,
and $\eta\pi^0$ channels from the corresponding cross sections measured by
the BELLE collaboration \cite{Nakazawa:2004gu,Chen:2006gy,Uehara:2009cf}.
Note that BELLE measured the cross section for the production of $K$-short
pairs, which is related to the $\KK$ cross section by
\begin{equation}
\sigma(\gamma\gamma\to\KK) = 2 \sigma(\gamma\gamma\to K_S K_S)\,.
\end{equation}
The c.m.\ energies we consider in our analysis include the charmonium
region.  For $\pi^0\pi^0$, $\pi^+\pi^-$, and $K^+K^-$ the BELLE
collaboration corrected their data for contributions from the $\chi_{cJ}$
states.  A corresponding subtraction was not performed for the $K_S K_S$
channel, where instead no cross section was given for $10.5 \gev^2 < s <
14.4 \gev^2$.  Note that $\chi_{cJ}$ states do not contribute to
$\eta\pi^0$ production.  The cross sections we calculate in the handbag
approach are of course to be understood as excluding contributions from
charmonium decays.

The extracted form factors are shown in the right panel of
Fig.~\ref{fig:FFpipi}.  As in the case of pion pairs, the $s$ dependence
of the form factors is somewhat steeper than the $1/s$ behavior predicted
by dimensional counting.  In particular, the $\KK$ form factor decreases
sharply with energy.  Such a behavior is not in conflict with the soft
handbag approach, given that the dimensional counting rules reflect the
dynamics of Brodsky-Lepage hard scattering rather than the dynamics of the
Feynman mechanism.  What the soft handbag approach \emph{does} predict is
that the non-valence form factor $R^s_{2\pi}$ should fall off more rapidly
with $s$ than its valence counterpart $R^u_{2\pi}$.  This is because the
soft mechanism requires that the momentum fraction of a final-state meson
with respect to the initial quark or antiquark must approach unity as $s$
increases \cite{Diehl:2001fv}.  In a pion this is of course more likely to
happen for $u\smash{\overline{u}}$ than for $s\smash{\overline{s}}$.

For $s\gsim 8\,\gev^2$ the $\KK$ form factor is smaller than predicted by
the relation
\begin{equation}
R_{\KK} \simeq \tfrac{2}{5} R_{K^+K^-} \,,
\label{eq:symm-rel}
\end{equation}
which is obtained from the SU(3) symmetry constraints
\eqref{old-form-factors} and \eqref{form-factors} if the non-valence form
factor $R^s_{2\pi}$ is negligible.  Thus, the behavior of $R_{\KK}$
clearly necessitates contributions from the non-valence form factor at
presently available energies.  The relation \eqref{eq:symm-rel} is then
expected to hold only for very large $s$.

\begin{figure}[ht]
\begin{center} 
\includegraphics[width=.38\tw,bb=148 287 477 710,clip=true]
{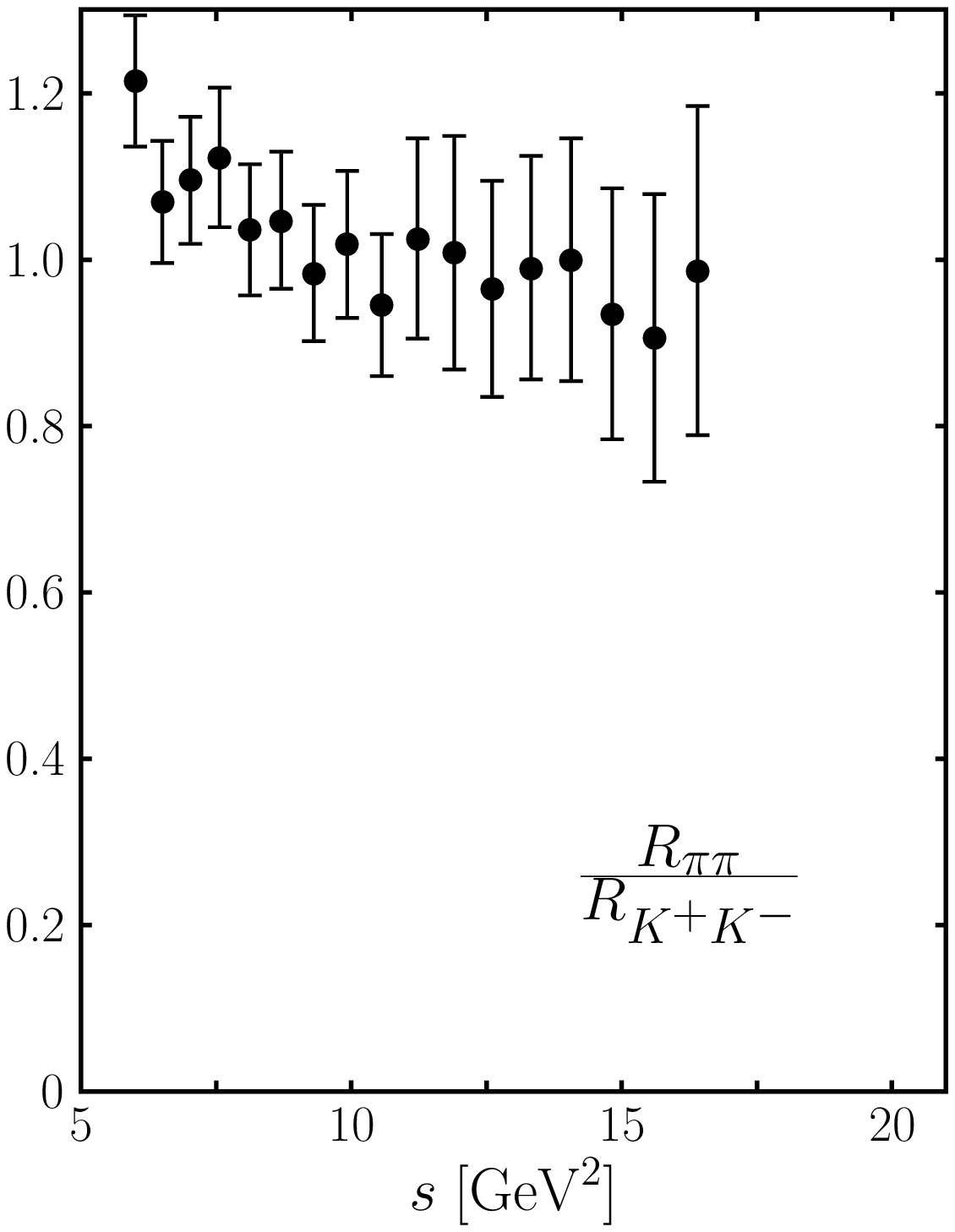}
\includegraphics[width=.43\tw,bb=150 269 524 700,clip=true]
{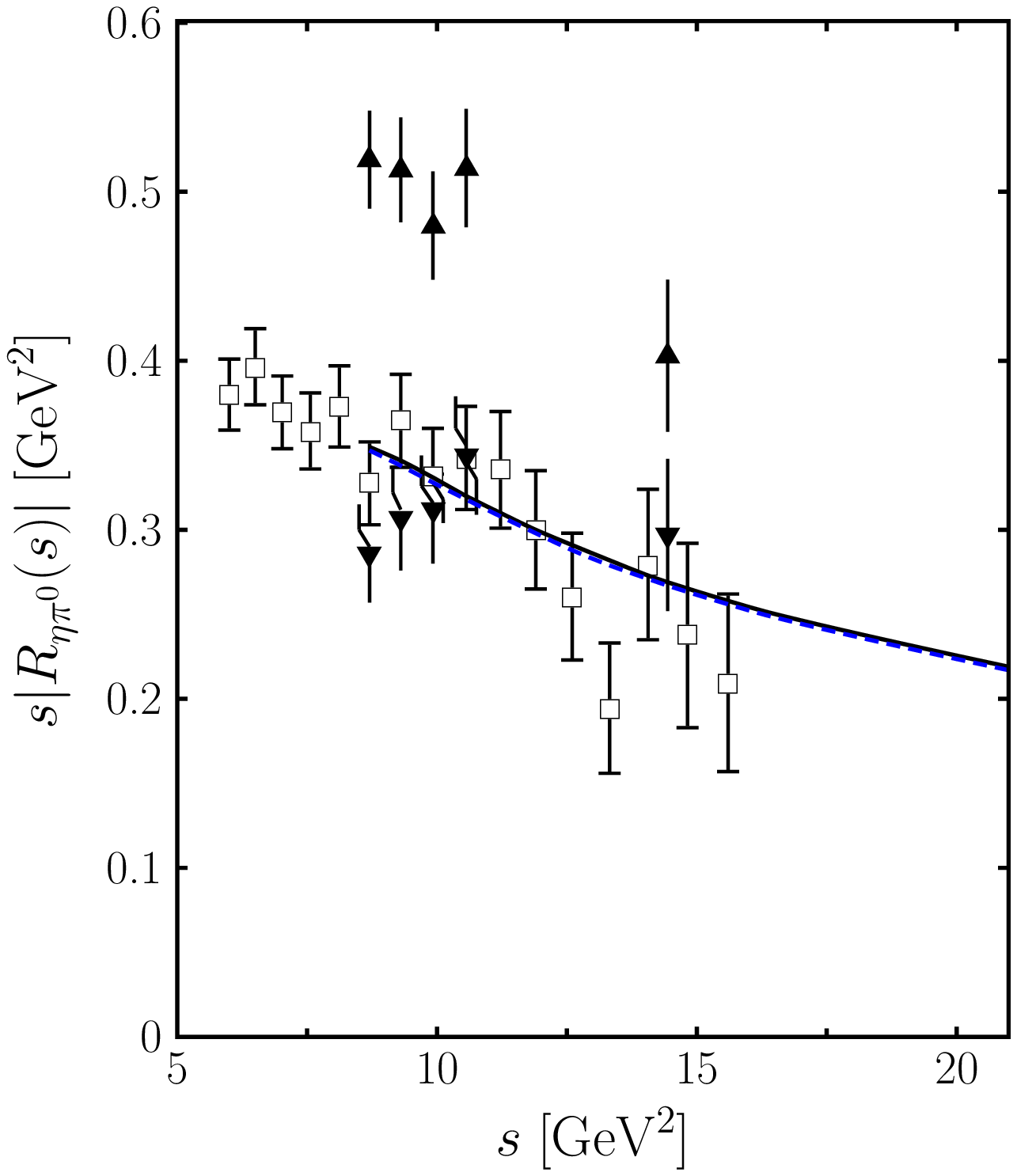}
\end{center}
\caption{\label{fig:FFratio} \emph{Left:} The ratio of $\pi\pi$ and
  $K^+K^-$ annihilation form factors, where $R_{\pi\pi}$ is obtained from
  the cross section combination \protect\eqref{eq:i=0}.  \emph{Right:}
  Probing the bounds \protect\eqref{upper-bound} and
  \protect\eqref{lower-bound} for the $\eta\pi^0$ form factor.  Open
  squares represent the scaled $\eta\pi^0$ form factor extracted from the
  data of \protect\cite{Uehara:2009cf}, whereas (reversed) triangles
  represent the upper (lower) bound evaluated from the data given in
  \protect\cite{Nakazawa:2004gu,Chen:2006gy}.  The solid and dashed curves
  have the same meaning as in Fig.~\protect\ref{fig:FFpipi}.}
\end{figure}

As quoted in \cite{Nakazawa:2004gu} the $U$-spin relation
\eqref{rel-SU3-1} between the $\pi^+\pi^-$ and $K^+K^-$ amplitudes (or
form factors) is fairly well respected by the data. The ratio of $K^+K^-$
and $\pi^+\pi^-$ cross sections is $0.89\pm 0.04 ({\rm stat.})\pm 0.15
({\rm syst.})$ for $s>9\,\gev^2$ \cite{Nakazawa:2004gu}.  When we use the
combination \eqref{eq:i=0} to extract $R_{\pi\pi}$ the agreement with
$R_{K^+K^-}$ is even better, as shown in the left panel of Fig.\
\ref{fig:FFratio}.  For $s\gsim 8\,\gev^2$ this ratio is perfectly
compatible with unity.  Even the errors on the ratio are not unreasonably
large.

As another test of the handbag approach combined with flavor symmetry, we
may verify the bounds \eqref{eq:bound} on the $\eta\pi^0$ cross section,
or equivalently on the corresponding form factor $R_{\eta\pi^0}$.  As is
evident from Fig.~\ref{fig:FFratio}, this form factor is much smaller than
the upper bound
\begin{equation}
  \label{upper-bound}
  \frac{|R_{K^+K^-}| + |R_{\KK}|}{\sqrt{3}}
\end{equation}
but rather close to  the lower bound 
\begin{equation}
  \label{lower-bound}
  \frac{|R_{K^+K^-}| - |R_{\KK}|}{\sqrt{3}} \,.
\end{equation}
For the first three energies above $s=9 \gev^2$, the lower bound is well
respected by the $\eta\pi^0$ data.  For $s=10.92\,\gev^2$ the bound is
practically saturated and for $s=14.44\,\gev^2$ it is slightly violated if
one considers the central values of the cross sections.  Within the
experimental errors, the lower bound is, however, consistent with the data
also at the highest energy.

\section{Combined analysis of meson octet channels}
\label{sec:combined}

\begin{figure}[t]
\begin{center} 
\includegraphics[width=.47\tw,bb=152 308 525 718,clip=true]
{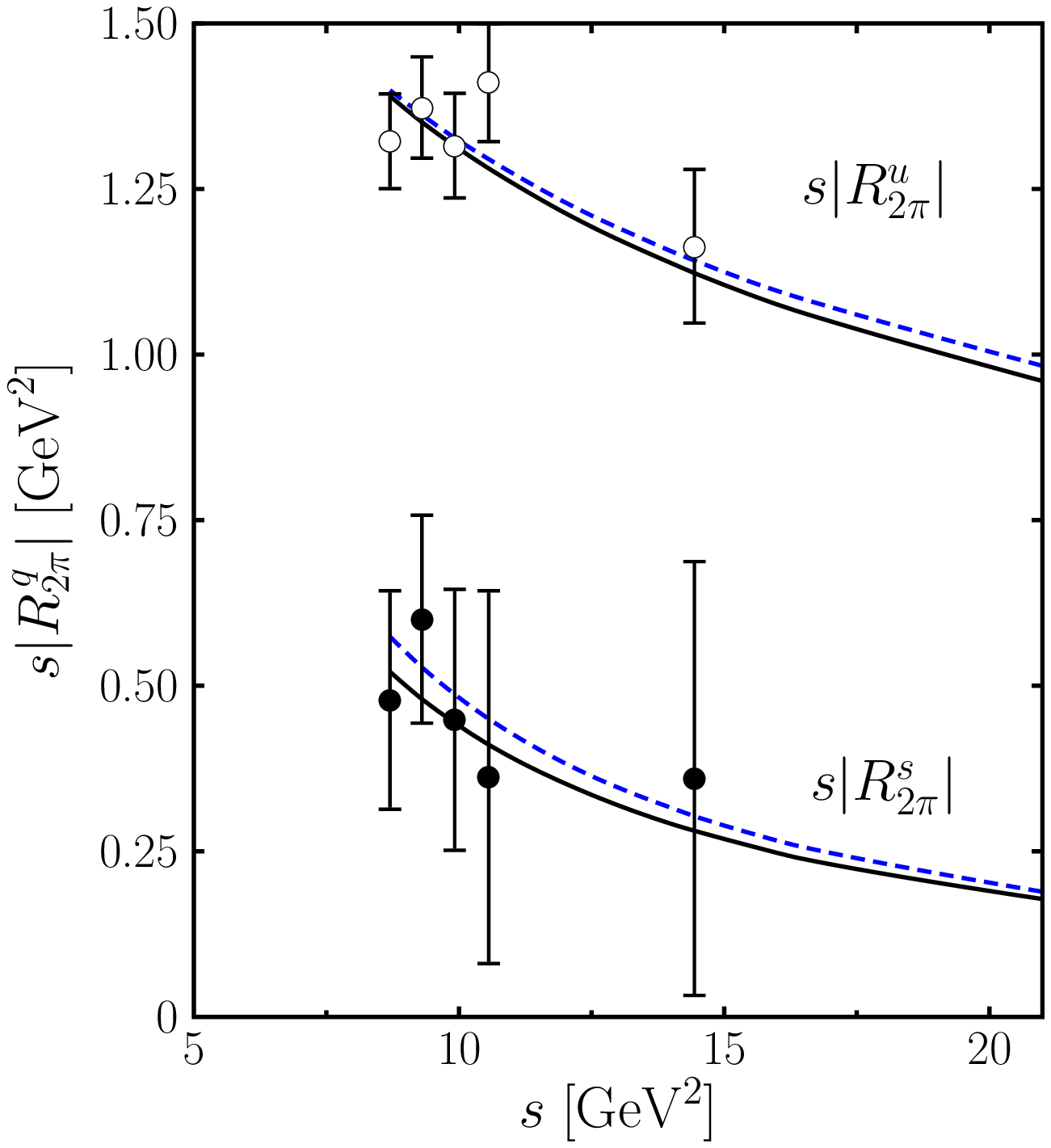}
\includegraphics[width=.435\tw,bb=181 308 525 718,clip=true]
{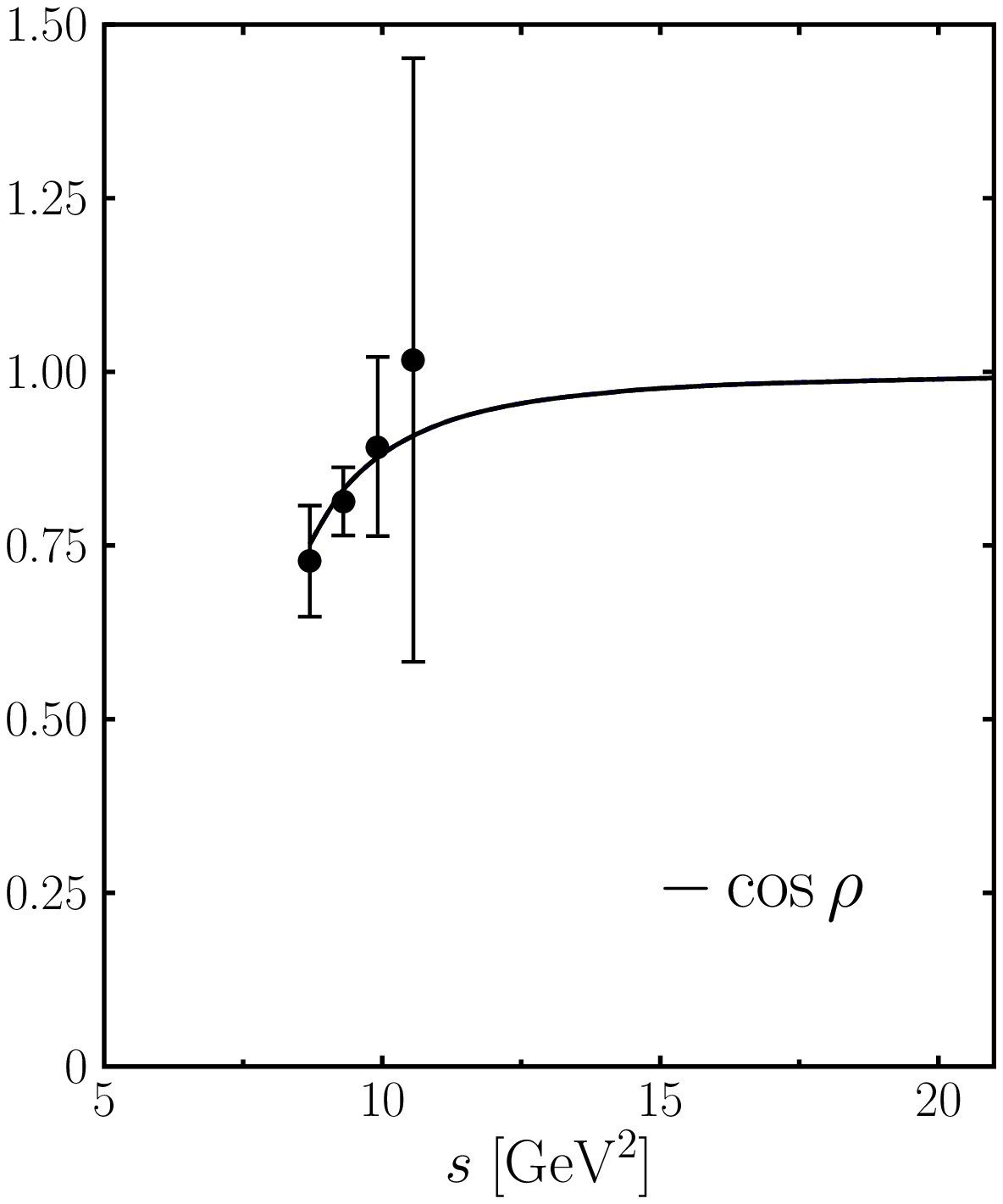}
\end{center}
\caption{\label{fig:basicFF} The scaled quark-level form factors
  $s|R^u_{2\pi}|$ and $s|R^s_{2\pi}|$ (left) as well as the cosine of
  their relative phase $\rho$ (right).  The data points are the results of
  the single-energy analysis described in the text.  The solid and dashed
  curves have the same meaning as in Fig.~\protect\ref{fig:FFpipi}.}
\end{figure}

Having investigated the annihilation form factors for the individual meson
channels, we now determine the basic valence and non-valence form factors
$R^u_{2\pi}$ and $R^s_{2\pi}$.  Given the good agreement between
$R^{}_{\pi\pi}$ and $R_{K^+K^-}$ documented in Fig.~\ref{fig:FFratio}, we
restrict our analysis to the data for $K^+K^-$, $\KK$, and $\eta\pi^0$
production from now on.  First we perform an analysis for each individual
value\footnote{%
At $s =14.44 \gev^2$ no results are given for the $K^+K^-$ and
$\eta\pi^0$ channels, and we interpolate by taking the weighted average of
cross sections at the two adjacent energies.}
of $s$ where the $\KK$ cross section is measured by BELLE
\cite{Chen:2006gy}.  We extract the quark-level form factors from the
relations in \eqref{eq:basic}, integrated over $\theta$ as specified in
Sect.~\ref{sec:1}.  The results of this procedure (henceforth referred to
as single-energy analysis) are listed in Tab.\ \ref{tab:1} and displayed
in the left panel of Fig.~\ref{fig:basicFF}.  While $R^u_{2\pi}$ is quite
well determined, $R^s_{2\pi}$ suffers from rather large uncertainties.
Nevertheless, a tendency for decreasing values of the scaled form factors
can be seen in the plot.  Particular noteworthy is the behavior of the
relative phase.  As shown in the right panel of Fig.~\ref{fig:basicFF},
the phase is strongly energy dependent and seems to tend towards
$180^\circ$.  In other words, the two form factors are found to be
essentially opposite in sign at the high-energy end of the data.

\begin{table*}[t]
\renewcommand{\arraystretch}{1.4} 
\begin{center}
\begin{tabular}{|c||c|c|c|}
\hline     
$s\, [\gev^2]$ &$\cos{\rho}$ &
   $s|R^u|\, [\gev^2]$ & $s|R^s|\, [\gev^2]$ \\[0.2em] \hline
\phantom{1}8.70	&$-0.73\pm 0.08$&$1.32\pm 0.07$	&$0.48\pm 0.17$	\\[0.2em]
\phantom{1}9.30	&$-0.81\pm 0.05$&$1.37\pm 0.08$	&$0.60\pm 0.16$ \\[0.2em]
\phantom{1}9.92	&$-0.89\pm 0.13$&$1.32\pm 0.08$	&$0.45\pm 0.20$	\\[0.2em]
10.56      	&$-1.02\pm 0.44$&$1.41\pm 0.09$ &$0.36\pm 0.28$	\\[0.2em]
14.44      	&$-1.00$        &$1.16\pm 0.12$	&$0.36\pm 0.33$	\\[0.2em]
14.44           &$-1.49\pm 3.80$&$1.16\pm 0.12$ &$0.19\pm 0.63$ \\[0.2em]
\hline  
\end{tabular}
\end{center}
\caption{\label{tab:1} Single-energy analysis of the data for $K^+K^-$,
  $\KK$, and $\eta\pi^0$.  Mixing between $\eta$ and $\eta^\prime$ is
  ignored.  The difference between the two entries for $s=14.44 \gev^2$ is
  explained in the text.}
\renewcommand{\arraystretch}{1.0}   
\end{table*}  

A remark concerning the data at $s=14.44\,\gev^2$ is in order.  It turns
out that at this energy the square of the non-valence form factor is
negative if the central values of the cross sections are inserted into
\eqref{eq:basic}.  There is, however, no inconsistency if the experimental
errors are taken into account.  Therefore, we determine the quark-level
form factors in the sense of a $\chi^2$-fit at this energy.  We utilize
two variants of the fit: in the first version we fix $\cos{\rho}$ at~$-1$
and fit only the absolute values of the two form factors. The
corresponding result is given in the penultimate row of Tab.\ \ref{tab:1},
and has $\chi^2_{\text{min}} = 0.55$ for three data points and two free
parameters.  In the second version of the fit, we allow $\cos{\rho}$ to
take any value.  This gives $\chi^2_{\text{min}} = 0.12$ and the values
listed in the last row of the table.  The errors on $\cos{\rho}$ and
$|R^s|$ are extremely large in this case.  The difficulties to determine
the quark-level form factors at $s=14.44\,\gev^2$ reflect the violation of
the lower bound for $R_{\eta\pi^0}$ by the central values of the cross
section, which we discussed at the end of the previous section.

So far we have neglected $\eta-\eta^\prime$ mixing in our analysis,
i.e.\ we have approximated the $\eta$ as a pure flavor octet meson
$\eta_8$.  To assess the uncertainties introduced by this approximation,
we repeat our single-energy analysis with an estimate of mixing effects.
For simplicity (and lack of better information) we assume that the
amplitude for annihilation into a pair of flavor-singlet mesons,
$\gamma\gamma \to \eta_1\eta_1$, is negligible.  We then only need to
multiply the $\eta\pi^0$ cross section obtained in the pure octet
approximation by $\cos^2{\theta_{PS}}$.  For the pseudoscalar mixing angle
$\theta_{PS}$ we take the value $15.4^\circ$ from \cite{Feldmann:1998vh},
which corresponds to $\cos^2{\theta_{PS}} = 0.93$.  We find that this
changes the results only marginally, the differences being well within the
errors quoted in Tab.~\ref{tab:1}.
     
In order to reduce errors and to exploit the full set of experimental
data, we now perform a combined energy-dependent fit to the three channels
$K^+K^-$, $\KK$, and $\eta\pi^0$.  For this fit we parameterize the form
factors as
\begin{align}
s |R^u_{2\pi}| &= a_u\ms \Bigl(\frac{s_0}{s}\Bigr)^{n_u} \,, &
s |R^s_{2\pi}| &= a_s\ms \Bigl(\frac{s_0}{s}\Bigr)^{n_s} \,,
\label{parameterization}
\end{align}
where we chose $s_0=9\,\gev^2$.  Guided by the results listed in
Tab.~\ref{tab:1} we take the following parameterization of the relative
phase:
\begin{equation}
\rho = \pi \biggl[\ms 1+ \tanh \frac{\kappa}{s-s_c} \ms\biggr] \,,
\end{equation}
where $\kappa$ is constant to be adjusted to the data.  For $s\to\infty$
we have $\rho\to \pi$.  The constant $s_c$ is introduced in order to have
a strong variation of $\cos{\rho}$ in the range of $s$ between $8.7$ and
$14.44\,\gev^2$, as is suggested by the results of the single-energy
analysis (see Fig.~\ref{fig:basicFF}).  Our default choice for this
parameter is $s_c=6.0\,\gev^2$, and as alternatives we take $s_c=5$ and
$7\,\gev^2$.  We also perform a fit with an estimate of $\eta -
\eta^\prime$ mixing, assuming again that $\eta_1\eta_1$ production is
negligible.  We fit all data with $s \ge 8.7\,\gev^2$ and give the
resulting parameters in Tab.~\ref{tab:2}.  With 28 data points altogether,
the minimum $\chi^2$ per degree of freedom is very low for all fits.  We
should recall in this context that we have included both the statistical
and systematic errors of the data when evaluating $\chi^2$.  As some of
the systematic uncertainties are not uncorrelated between the data points,
a more refined analysis would give larger values of $\chi^2_{\text{min}}$
than those in our table.

\begin{table*}[t]
\renewcommand{\arraystretch}{1.4} 
\begin{center}
\begin{tabular}{|c|c|c||l|l|l|l|l|}
\hline     
 fit &  $s_c\, [\gev^2]$ & ~$\chi^2_{\text{min}}$~  &  
  \multicolumn{1}{c|}{$a_u\, [\gev^2]$} &
  \multicolumn{1}{c|}{$n_u$} &
  \multicolumn{1}{c|}{$a_s\, [\gev^2]$} &
  \multicolumn{1}{c|}{$n_s$} &
  \multicolumn{1}{c|}{$\kappa\, [\gev^2]$} \\[0.2em]
\hline 
 1  & $6.0$ & $14.6$ & $1.37\pm 0.03$ & $0.42\pm 0.08$
    & $0.50\pm 0.05$ & $1.22\pm 0.40$ & $0.63\pm 0.06$ \\[0.2em]
 2  & $6.0$ & $13.8$ & $1.38$ & $0.40$ & $0.55$ & $1.26$ & $0.63$ \\[0.2em]
 3  & $7.0$ & $14.8$ & $1.38$ & $0.43$ & $0.46$ & $1.09$ & $0.40$ \\[0.2em]  
 4  & $5.0$ & $14.9$ & $1.37$ & $0.42$ & $0.50$ & $1.21$ & $0.84$ \\[0.2em]
\hline 
\end{tabular}
\end{center}
\caption{\label{tab:2} Energy-dependent fits to the $K^+K^-$, $\KK$, and
  $\eta\pi^0$ data with $s \ge 8.7\,\gev^2$. Fit 2 includes an estimate of
  $\eta - \eta^\prime$ mixing as explained in the text, whereas fits 1, 3,
  and 4 neglect mixing.  The errors on the parameters in fits 2 to 4 are
  very similar to those in fit 1 and not listed for better legibility.}
\renewcommand{\arraystretch}{1.0}   
\end{table*} 

In Fig.~\ref{fig:FFpipi} we compare the process form factors evaluated
from fits 1 and 2 with those we directly extracted from the BELLE data.
The results of the two fits practically coincide and agree well with the
data.  We also compare the fit results with the quark-level form factors
we derived in our single-energy analysis.  Again good agreement is
obtained, as shown in Fig.~\ref{fig:basicFF}.  We observe that the fitted
form factors \eqref{parameterization} do not respect dimensional counting,
and that the non-valence form factor falls off more rapidly with energy,
so that at large energies the valence form factor is dominant.  As
discussed in the previous section, this is fully consistent with the soft
handbag picture.  However, due to the weight factors of $R_{2\pi}^{u}$ and
$R_{2\pi}^s$ in \eqref{form-factors}, the approach to valence dominance is
very slow for $R_{\KK}$.  At $s=100\,\gev^2$, for instance, the valence
dominance relation \eqref{eq:symm-rel} still receives corrections of about
$10\%$ when evaluated with our fitted form factors.  In contrast to
$R_{\KK}$, the form factor $R_{K^+K^-}$ is strongly dominated by the
valence contribution.  This provides some justification for the neglect of
the non-valence form factor in our previous analysis \cite{Diehl:2001fv},
where we had only data for the $\pi^+\pi^-$ and $K^+K^-$ channels at our
disposal.

\begin{figure}[ht]
\begin{center} 
\includegraphics[width=.48\tw,bb=146 255 523 700,clip=true]
{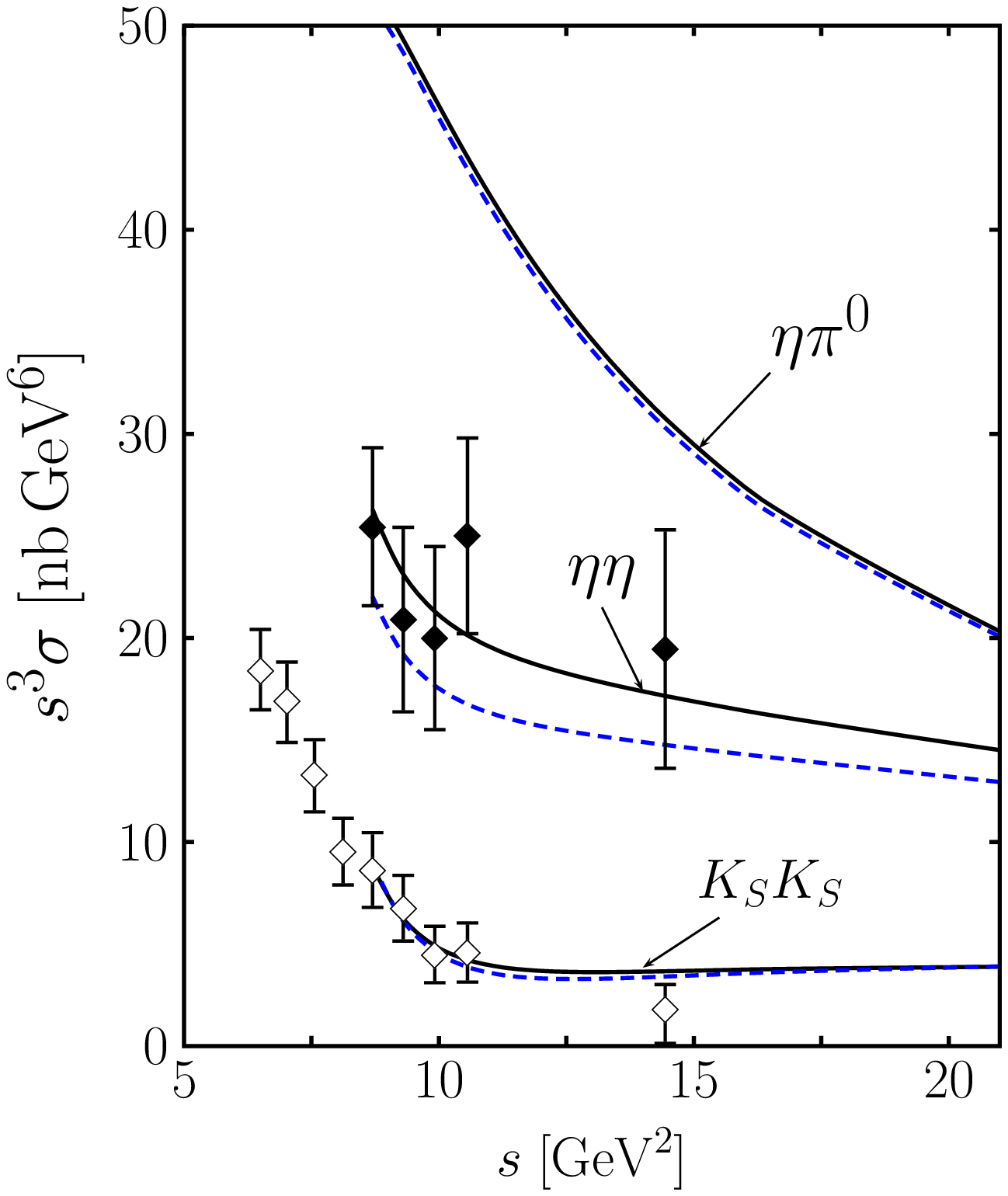}
\end{center}
\caption{\label{fig:cross} Predictions for the $\eta\eta$ cross section,
  integrated over $-0.6 < \cos\theta < 0.6$ and scaled by $s^3$.  For
  comparison we also show the corresponding $\eta\pi^0$ and the
  $K_S K_S$ cross sections, with the kaon data (open diamonds)
  taken from \protect\cite{Chen:2006gy}.  The filled diamonds represent
  the $\eta\eta$ cross section evaluated from the BELLE data
  \protect\cite{Nakazawa:2004gu,Chen:2006gy,Uehara:2009cf} with the
  relation \protect\eqref{eq:cross-section-relation}.  The solid (dashed)
  curves represent the cross sections obtained from fit 1 (2) in
  Tab.~\protect\ref{tab:2}; note that they do not include contributions
  from charmonium decays.}
\end{figure}

Figure~\ref{fig:cross} finally shows the integrated cross sections for
$\eta\pi^0$, $\eta\eta$, and $K_S K_S$ production resulting from our fits.
The $\eta\eta$ cross section has not been measured as yet.  We also
estimate it directly from the $K^+K^-$, $\KK$, and $\eta\pi^0$ cross
sections using the SU(3) relation~\eqref{eq:cross-section-relation}, and
we find good agreement with the fit results.  In contrast to all other
quantities discussed so far, $\eta - \eta^\prime$ mixing is of greater
importance for the $\eta\eta$ cross section, which is multiplied by
$\cos^4{\theta_{PS}} \approx 0.86$ in fit 2.  The difference between fits
1 and 2 can be considered as the uncertainty of our prediction for this
cross section due to $\eta - \eta^\prime$ mixing.

The ALEPH collaboration has also measured the cross sections for
two-photon annihilation into $\pi^+\pi^-$ and $K^+K^-$
\cite{Heister:2003ae}.  The errors of these data are, however, so large
that they have no influence if we include them in our fits.  Of some
interest are the two data points for the $\pi^+\pi^-$ channel at energies
larger than those of the BELLE results.  From these data we obtain
\begin{align}
s |R_{\pi\pi}| &= 0.84\pm 0.42 \gev^2 & \text{at}~ s &= 18.1 \gev^2 \,,
\nonumber \\ 
               &= 0.80\pm 0.34 \gev^2 &              &= 27.6 \gev^2
\end{align}
when neglecting a possible $I=2$ contribution, which we expect to be small
at these high energies.  The values of the form factor from our fit 1 are
$s |R_{\pi\pi}| = 0.59 \gev^2$ at $s= 18.1 \gev^2$ and $s |R_{\pi\pi}| =
0.48 \gev^2$ at $s= 27.6 \gev^2$, in fair agreement with the ALEPH data.

\section{Some remarks on the hard-scattering mechanism}
\label{sec:brodsky-lepage}

In this section we point out some differences between the soft handbag
approach and the hard-scattering mechanism of Brodsky and Lepage
\cite{Lepage:1980fj}.  Given the focus of the present work, we emphasize
the relations between different meson channels in the two mechanisms.

The nonperturbative quantities required in the hard-scattering picture are
the leading-twist meson distribution amplitudes, characterized by the
meson decay constants and the normalized distribution amplitudes
$\phi_M(z)$.  The asymptotic shape $\phi_M(z) = 6z(1-z)$ has been used for
many channels in \cite{Brodsky:1981rp} and for $\pi^+\pi^-$ and $K^+K^-$
in \cite{Duplancic:2006nv}, where the hard-scattering subprocess was
evaluated at next-to-leading order in $\alpha_s$.  The relation between
the cross sections for different channels is then governed by charge
factors and by the meson decay constants, which are well known for the
pseudoscalar octet.  In particular the ratio of $K^+K^-$ and $\pi^+\pi^-$
cross sections is given by $(f_K/f_\pi)^4 \approx 2.0$ in this case, with
$f_K/f_\pi$ from \cite{Amsler:2008zzb}.  Such a strong breaking of the
SU(3) symmetry relation \eqref{rel-SU3-1} is ruled out by the BELLE data
\cite{Nakazawa:2004gu} for energies up to $s = 16.4 \gev^2$.  We note that
the absolute size of cross sections in the above calculations is
significantly below the data \cite{Duplancic:2006nv}.  This also holds in
the modified hard-scattering approach, where Sudakov resummation and
intrinsic transverse momentum of the quarks in the mesons are taken into
account \cite{Vogt:2000bz}.

At moderately large scales it is natural to expect that $\phi_M(z)$
deviates from its asymptotic form under evolution, and one can thus have
SU(3) symmetry breaking in the shapes of the distribution amplitudes.
This was investigated in \cite{Benayoun:1989ng,Chernyak:2006dk} using QCD
sum rule estimates.  With these estimates, the SU(3) breaking effects in
the shapes of distribution amplitudes largely compensate those in the
meson decay constants.  A ratio of $\pi^+\pi^-$ and $K^+K^-$ cross
sections close to unity was obtained, in agreement with the BELLE data.

The production of neutral meson pairs in the hard-scattering mechanism is
generically suppressed, since it turns out that at leading order in
$\alpha_s$ the bulk of the amplitude is proportional to a charge factor
$(e_{q_1} - e_{q_2})^2$ for a meson with quark content $q^{}_1
\overline{q}_2$ \cite{Brodsky:1981rp}.  The explicit calculations in
\cite{Brodsky:1981rp} and in \cite{Benayoun:1989ng,Chernyak:2006dk} yield
values below $0.05$ for the ratio $R$ of integrated $\pi^0\pi^0$ and
$\pi^+\pi^-$ cross sections.  According to \eqref{eq:isospin-decomp} this
requires $I=0$ and $I=2$ transitions of comparable size, in sharp contrast
to the situation in the handbag approach.  The values of $R$ obtained in
\cite{Brodsky:1981rp,Benayoun:1989ng,Chernyak:2006dk} are significantly
below the experimental results (shown in our Fig.~\ref{fig:cross-ratio}).
Because of charge factors, the ratio of $K_S K_S$ and $K^+K^-$ cross
sections in \cite{Benayoun:1989ng,Chernyak:2006dk} is even smaller, and it
was concluded in \cite{Chernyak:2006dk} that at BELLE energies the $K_S
K_S$ channel is dominated by contributions other than the hard-scattering
one.

As for the $\eta\eta$ channel, our prediction for the $\eta\eta$ cross
section differs markedly from the one obtained with asymptotic
distribution amplitudes in \cite{Brodsky:1981rp}, where the ratio of
$\eta\eta$ and $\pi^0\pi^0$ cross sections is given by the factor $0.4\ms
(f_\eta/f_\pi)^4$.  If one ignores $\eta - \eta^\prime$ mixing and takes
for the ratio of $\eta$ and $\pi$ decay constants the value $1.28$ from
\cite{Feldmann:1998vh}, one obtains a cross section ratio of about $1.1$,
whereas our fit 1 in the handbag approach gives about $0.3$.

We conclude this section with the remark that the new BaBar measurement
\cite{Aubert:2009mc} of the $\pi\gamma$ transition form factor at high
momentum transfer is challenging our knowledge about the shape of the pion
distribution amplitude.  In this light one may expect that numerical
estimates for $\gamma\gamma\to\MM$ within the hard-scattering picture will
need to be revised.

\section{Summary}
\label{sec:sum}

Combined with SU(3) flavor symmetry, the soft handbag approach for
two-photon annihilation into pairs of pseudoscalar mesons provides a good
description of the experimental data for $s\gsim 9 \gev^2$, which have
been obtained by the BELLE collaboration in the recent years.  The
measured ratio of $\pi^0\pi^0$ and $\pi^+\pi^-$ production rates implies
the presence of an isospin $I=2$ contribution at $s\sim 9 \gev^2$, which
cannot be generated by the handbag mechanism.  Depending on its relative
phase, this contribution can, however, be as small as 10\% at the
amplitude level, which we regard as a quite tolerable correction to the
handbag approach at this energy.

Flavor symmetry allows one to express the annihilation form factors for
all processes in terms of only two quark-level form factors, one for
valence quarks and a second one for non-valence quarks.  We have
determined these form factors in two ways, first in a single-energy
analysis and then in energy-dependent fits, with good agreement between
the two methods.  In our energy-dependent fits we find that the
non-valence form factor is suppressed by nearly a power of $s$ compared to
the valence one.  This is in agreement with the physics of the handbag,
which requires that the quark or antiquark entering the meson take most of
its momentum.  We find that the relative phase between the two form
factors tends to $180^\circ$ at the highest energies ($s\sim 16 \gev^2$)
of the data.

The success of our analysis implies that the present data do not provide
evidence for SU(3) flavor violation in the annihilation form factors at
the level of the experimental errors, which are typically less than $10 -
15\%$.  The ultimate confirmation of this picture would be an experimental
verification of our prediction for the $\gamma\gamma\to\eta\eta$ cross
section.

\section*{Acknowledgments} 

We gratefully acknowledge correspondence with S.~Uehara.  This work is
supported in part by BMBF, contract no.~06RY258, and by the European
Project Hadron Physic 2IA in EU FP7.


\end{document}